\documentclass[paper=a4,abstracton,numbers=noenddot]{scrartcl}
\usepackage[a4paper,left=2.5cm,right=2.5cm,top=3cm,bottom=2.5cm]{geometry}

\usepackage{authblk}
\usepackage[english]{babel}  
\usepackage{hyperref}
\usepackage[intlimits]{amsmath}
\usepackage{amsthm}
\usepackage{amssymb} 
\usepackage{units} 
\usepackage{amsfonts,bbm}
\allowdisplaybreaks[1]
\usepackage[numbers]{natbib}
\usepackage{booktabs} 
\usepackage{color}
\definecolor{myblue}{rgb}{0,0,0.68} 
\definecolor{myorange}{rgb}{1,0.5,0}
\newcommand{\km}[1]{#1}

\newcommand{\mb}[1]{#1}
\usepackage{caption}
\usepackage{graphicx}
\graphicspath{{.}{./pics/}{./picsWS2022}{./pics/picsWS2022}}
\usepackage{cleveref}
\crefformat{figure}{Fig.~#2#1#3} 
\Crefformat{figure}{Fig.~#2#1#3} 
\crefformat{table}{Table~#2#1#3} 
\Crefformat{table}{Table~#2#1#3} 
\crefformat{equation}{(#2#1#3)} 
\Crefformat{equation}{(#2#1#3)} 
\labelcrefformat{equation}{(#2#1#3)}  
\crefformat{section}{Section~#2#1#3} 
\Crefformat{section}{Section~#2#1#3} 

\title{Teaching mechanics with individual exercise assignments and automated correction}

\begin{document}
\author[1]{M.~H.~Gfrerer}
\author[1]{B.~Marussig}%
\author[2]{K.~Maitz}
\author[3]{M.~M.~Bangerl}
\affil[1]{Institute of Applied Mechanics, Graz University of Technology}
\affil[2]{Private University College of Teacher Education Augustinum (PPH Augustinum)}
\affil[3]{Institute of Interactive Systems and Data Science, Graz University of Technology}

\maketitle                   

\begin{abstract}
    Solving exercise problems by yourself is a vital part of developing a mechanical understanding. Yet, most mechanics lectures have more than 200 participants, so the workload for manually creating and correcting assignments limits the number of exercises. The resulting example pool is usually much smaller than the number of participants, making verifying whether students can solve problems themselves considerably harder. At the same time, unreflected copying of tasks already solved does not foster the understanding of the subject and leads to a false self-assessment. 
    We address these issues by providing a scalable approach for creating, distributing, and correcting exercise assignments for problems related to statics, strength of materials, dynamics, and hydrostatics. The overall concept allows us to provide individual exercise assignments for each student. 
    A quantitative survey among students of our recent statics lecture assesses the acceptance of our teaching tool. 
    The feedback indicates a clear added value for the lecture, which fosters self-directed and reflective learning.
\end{abstract}

\section{Motivation}

Lecturers in mechanics are often confronted with introductory bachelor courses with several hundred students. Moreover, the students' prior knowledge of the subjects, such as statics, strength of materials, and dynamics, is very heterogeneous. 
The ultimate goal is to support all students to understand the fundamental concepts in such a depth that allows them to solve complex engineering problems.
To achieve such a level of understanding, the students' active application of the teaching content is crucial since it helps to reorganize and transfer the information provided in the lecture hall. 

Homework assignments are a classic tool to foster the active use of the content taught, but there are difficulties, especially for courses with hundreds of students.
From a lecturer's perspective, the number of assignments is usually limited by the amount of work involved in creating and correcting them.
Consequently, often only a small pool of different examples is provided, opening the possibility of a rapid exchange of supposed sample solutions between students. 
While student discussions on how to solve a problem are very valuable, unreflected copying of already solved assignments does not promote the understanding of the subject, results in unfair grading, and leads to a false self-assessment of the student's knowledge.

We aim to mitigate these undesired aspects of homework assignments by proposing a concept that \emph{automatically} creates, distributes, and corrects example problems. Hence, we can provide \emph{individual exercise assignments} for each student. 
These scalable individual assignments decouple a lecturer's workload from the number of students and allow a better assessment of the student's knowledge. Furthermore, it provides the possibility to adjust the complexity to the student's background. 
Besides these advantages from the lecturers' perspective, our solution aims to support students in their learning process. The automatic correction yields more immediate and tailored feedback to student answers, the possibility to resubmit assignments enables reflective learning, and comparing colleagues' solutions potentially results in discussion rather than a copying of calculations.
However, in order for such a learning technology to be successfully adopted by students and the potential advantages to become effective, the technology needs to be easy to use and usable~\cite{granic2019technology} and -- amongst other factors related to individual attitude and motivation -- the potential benefits for their own learning must be clear to the students~\cite{kemp2019taxonomy}.

Therefore, the aim of this paper is twofold: First, we want to introduce our solution and provide a description of the technological features and implementation. Second, we analyze how the solution is perceived by the students using it and present the results of the first implementation and evaluation of the individual exercise assignment concept in the field.

\section{Individual exercise assignments}

Our solution consists of two parts: the back-end for an automatic generation of the actual assignments and the front-end that provides the user interface for the students.
\Cref{sec:implementationFeatures} highlights specific features of our tool, while \cref{sec:techincalImplementation} discusses not yet published technical implementation aspects and provides further references.

\subsection{Implementation features}
\label{sec:implementationFeatures}
\subsubsection{Back-end}
The back-end system ``mechpy'' described here is implemented in Python and encompasses several features. 
It allows the generation of similar assignments and their solutions. 
Furthermore, ``mechpy'' creates the assignment instructions and formulates the rules for correcting student answers so that they are ready for the front-end system presented later in \cref{sec:frontend}. 

One essential capability of ``mechpy'' is that all calculations can be done symbolically.
Thus, it is possible to perform calculations with variables and symbols instead of just numerical values. 
This feature is vital since it allows the creation of numerical reference solutions that are identical to the sought handwritten results.

\begin{figure}[b]
	\begin{minipage}[b]{0.49\textwidth}
        \centering 
        \includegraphics[width=0.75\linewidth,trim={0 0.4cm 0cm 4.8cm},clip]{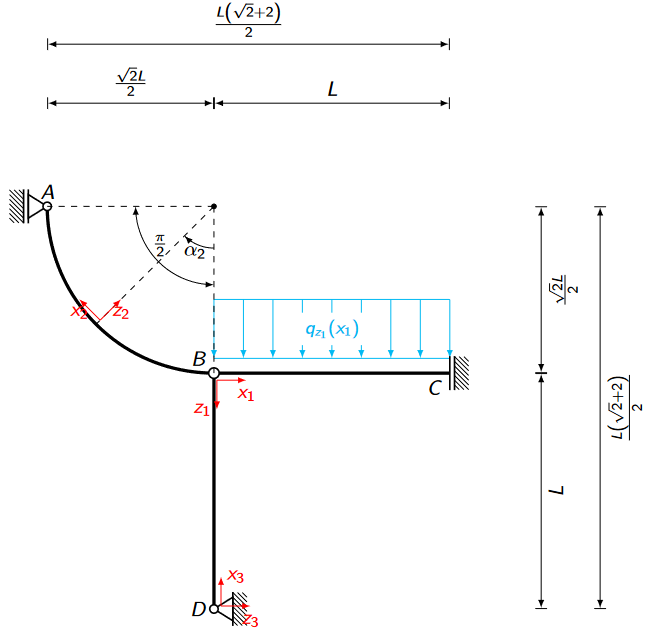}\\
        {a) Mechanical system }
    \end{minipage}    
	\begin{minipage}[b]{0.49\textwidth}
        \centering 
        \includegraphics[width=0.9\linewidth,trim={0 1.5cm 0cm 3.5cm},clip]{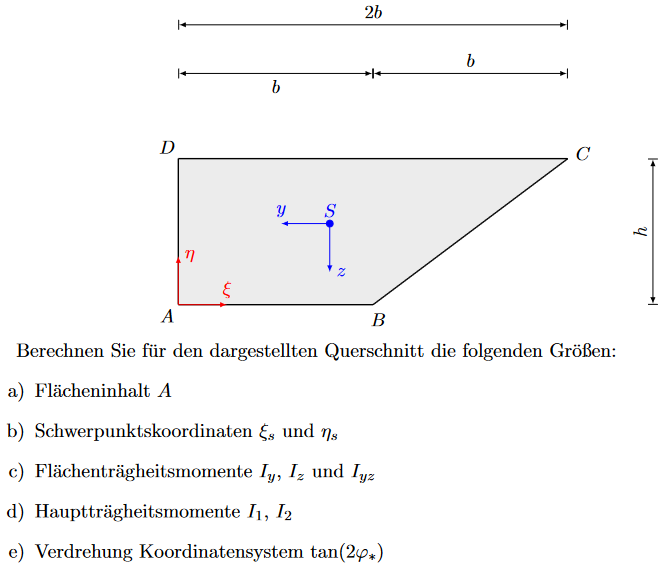}\\
        {b) Cross-section}
     \end{minipage} 
    \caption{The main problem types supported by the back-end. Note that example values are provided as parameters and not as numbers.}
    \label{fig:ProblemTypes}     
\end{figure}
\begin{figure}[t]
    \begin{captionbeside}{System consisting of flexible beams ($\overline{BA}$,$\overline{BC}$ with $EI$ and $EA$) and a rigid beam ($\overline{DB}$ with $EI\rightarrow\infty$ and $EA\rightarrow\infty$).}
        [r]
        \centering 
        \begin{minipage}[b]{0.76\textwidth}		
        \begin{minipage}[b]{0.475\textwidth}		
            \includegraphics[width=0.9\linewidth,trim=5cm 6.9cm 8cm 14.9cm,clip]{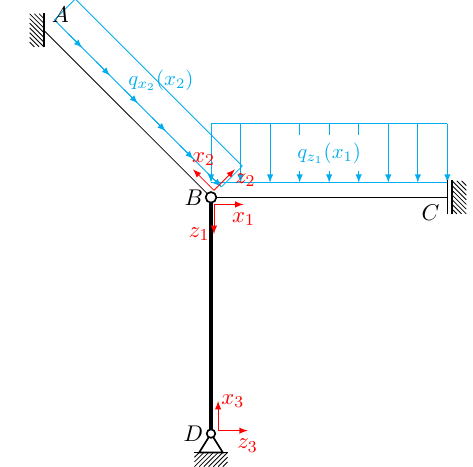}\\%
            {a) Example description}
        \end{minipage}\hfill
        \begin{minipage}[b]{0.475\textwidth}
            \vspace{-1.5cm}
            \includegraphics[width=0.85\linewidth]{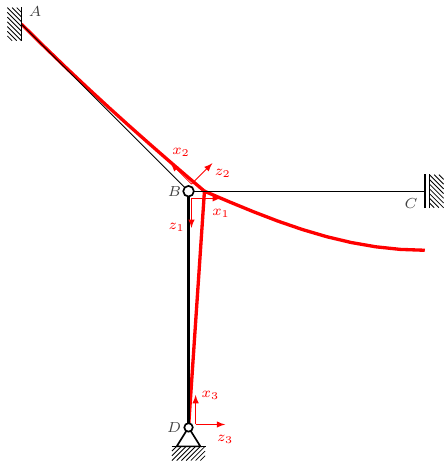}\\
            {b) Deformed shape}
        \end{minipage}
        \end{minipage}
    \end{captionbeside}
    \label{fig:rigid}
\end{figure}
\begin{figure}[b]
	\begin{minipage}[t]{0.23\textwidth}
        \centering
        \includegraphics[scale=0.33,trim={0 0 3cm 1.7cm},clip]{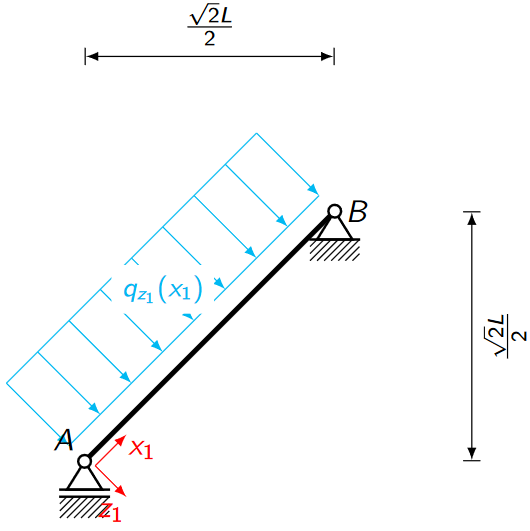}\\
        {a) fully local load}
	\end{minipage}\hfill
	\begin{minipage}[t]{0.23\textwidth}
        \centering
        \includegraphics[scale=0.334,trim={0 0 3cm 2.2cm},clip]{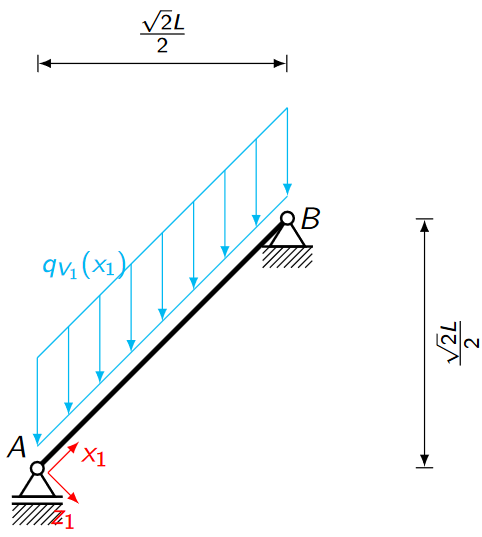}\\
        {b) local-global load}
	\end{minipage}
	\begin{minipage}[t]{0.23\textwidth}
        \centering
        \includegraphics[scale=0.33,trim={0 0 3.8cm 2cm},clip]{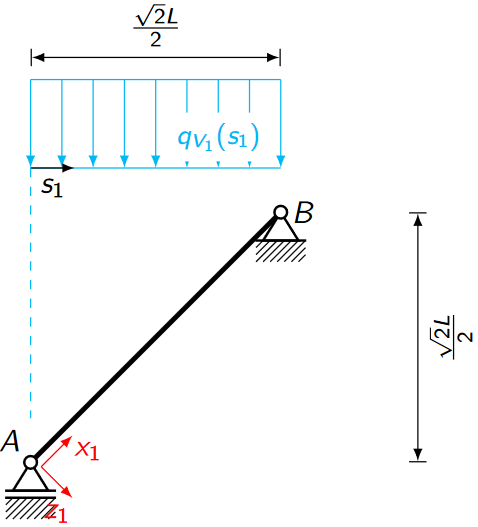}\\
        {c) fully global load}
	\end{minipage}    
	\begin{minipage}[t]{0.24\textwidth}
        \centering
        \includegraphics[width=1.0\linewidth,trim={0 0 4cm 3cm},clip]{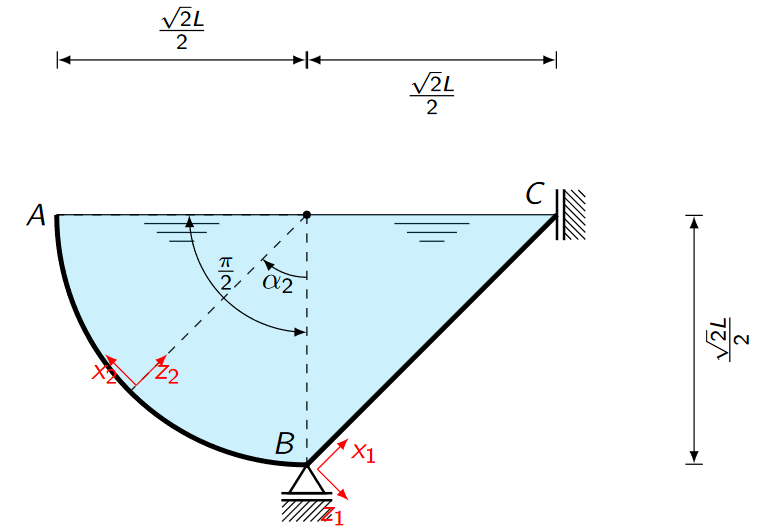}\\
        {d) hydrostatic load}
	\end{minipage}
 \caption{Implemented distributed load types: (a) load direction and magnitude defined w.r.t. the local coordinates of the beam, (b) load direction defined w.r.t. the global coordinates, whereas the magnitude is a function of the local position x1, (c) load direction and magnitude defined w.r.t. the global coordinates, and (d) hydrostatic loading}
 \label{fig:Loads}
\end{figure}

The two considered problem classes are ``mechanical system'' and ``cross-section'' examples, see \cref{fig:ProblemTypes}. 
For mechanical systems, the analysis of statically determined and statically undetermined systems is possible. 
The former may consist of straight beams and circular arcs, while the latter is currently restricted to straight beams only since this is also the focus of our corresponding lectures. In any case, a notable feature is the possibility to pass to the limits of axial rigid beams ($EA\rightarrow\infty$) and/or flexural rigid beams ($EI\rightarrow\infty$), as shown in Fig.~\ref{fig:rigid}.
\Cref{fig:Loads}, on the other hand, summarizes the currently supported loading types.
Regarding the ``cross-section'' examples, the back-end utilizes the divergence theorem to compute all quantities of interest (e.g., area, center of gravity, and moment of inertia of area). As discussed later in \cref{sec:techincalImplementation}, this theorem relates the flux of a vector field through a closed curve to the divergence of the field within the volume enclosed by the curve. Thus, only the integration along the input boundary curve, which defines the cross-section's shape, is necessary.

With the above-described capabilities at hand, we can compute symbolic reference solutions of an assignment example. To generate similar assignments, ``mechpy'' alters  problem parameters by user-defined assembly rules. For cross-section examples, the boundary curves are varied, while the parameters for mechanical systems include 
\begin{itemize}
    \item orientation and position of supports, straight beams, and circular arcs, 
    \item as well as joint and loading types and positions.
\end{itemize}
Finally, the resulting examples are encoded in LaTeX to compile the assignment visualizations and instructions.

Overall, ``mechpy'' generates (i) a large number of similar problems and their instructions, (ii) the corresponding symbolic solutions, and (iii) the rules for the correction of student answers. All this information is exported as a XML file compatible with the front-end application.

\subsubsection{Front-end}
\label{sec:frontend}

We employ the learning management system Moodle\footnote{\url{https://moodle.org/}, accessed 30.6.2023} as the front-end. 
When importing the XML file generated by the back-end ``mechpy'', the exercise assignments are available as a question pool.
Hence, the assignment can be used just like any other question in Moodle. 
In particular, we use following the features:
\begin{itemize}
    \item Random distributions of a question pool's assignments among students
    \item Automatic submission of student answers with predefined rules (e.g., deadlines, number of re-submissions, etc.)
    \item Automatic interpretation and grading of student answers
\end{itemize}

To realize the last point, the ``mechpy'' assignments are formulated as Moodle STACK\footnote{\url{https://moodle.org/plugins/qtype_stack}, accessed 30.6.2023} questions.
STACK is an open-source automatic assessment system based on the computer algebra system Maxima\footnote{\url{https://maxima.sourceforge.io/}, accessed 30.6.2023}, which allows the interpretation and modification of mathematical expressions.
Due to its vast scope, STACK is more complex than simple multiple-choice questions, yet, the interface is simple.
\begin{figure}[t]
    \centering 
    \includegraphics[width=1.0\textwidth]{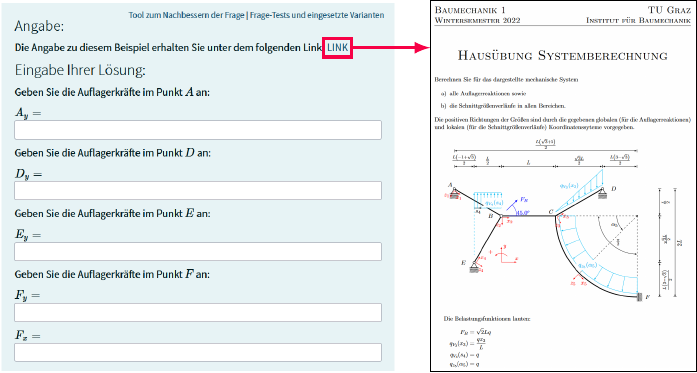}
    \caption{Graphical user interface of the front-end: (left) fields for the student input and (right) the instructions provided as PDF file.}
    \label{fig:Interface}     
\end{figure}
As shown in \cref{fig:Interface}, the instructions are provided by a linked PDF file, which is also encoded in the XML import from the back-end.
Furthermore, the answers can be written directly into the web interface using different syntaxes (e.g., $x^2:=$`x**2' or `x$\wedge$2') and students get direct feedback on how STACK interprets their input. Hence, they can check for typos before the answer is actually compared with the reference solution. 

STACK provides a wide range of tests to compare student answers with reference solutions. The default test evaluates if two expressions are algebraically equivalent. Since the ``mechpy'' reference solutions are symbolic, we could use this approach. However, the expression may become lengthy for some assignment examples. Hence, we decided that students can provide their answers also with floating point numbers to increase the user-friendliness of the tool.
As a result, we must employ the numerical tests of STACK for the correction. These tests include consideration regarding the accuracy of the input. Nevertheless, situations where the reference solution (or parts of it) become almost zero, require special attention. Therefore, ``mechpy'' introduces several STACK-subtests before assessing the student answers.   
Despite the numerical test, using a symbolic reference solution is still beneficial since only the input introduces the accuracy errors. For instance, answers rounded not at the end of the calculation but somewhere in between are usually detected as wrong, especially when the rounding is done in the beginning, as it would be the case when the examples are imported into a numerical solutions tool, e.g., a 
CAD software for computing cross-section quantities.

\subsection{Technical implementation}
\label{sec:techincalImplementation}
The technical implementation for statically determined systems can be found in \cite{Kramer2022,Krenn2022BaThesis}, whereas \cite{SoellradlBaThesis} details the treatment of statically undetermined systems and axial/flexural rigid beams. 
Thus, here we focus on the solution procedure for cross-section examples.

In our cross-section examples, the first task is always to compute the cross-section area. While the students solve this task by subdividing the cross-section into elementary shapes, like triangles and quadrilaterals, we use the divergence theorem in the back-end.  
Let $\mathbf V$ be a vector field defined on the region $A$ with boundary $\partial A$ and outward normal vector $\mathbf n$. Then, the divergence theorem states that the area integral of the divergence of $\mathbf V$ can be expressed as an integral over the boundary
\begin{equation}\label{eq::divergenceTheorem}
    \int_A \operatorname{div}\mathbf V \;dx = \int_{\partial A} \mathbf V \cdot \mathbf n \;ds.
\end{equation}
To use \eqref{eq::divergenceTheorem} for the calculation of the cross-section area, we observe that for $V = \tfrac{1}{2}[x,y]^\top$, we have $\operatorname{div}V = 1$, and thus
\begin{equation}
  \int_A 1\; dx = \int_A\frac{1}{2}\operatorname{div}\begin{bmatrix}
      x\\y
  \end{bmatrix} dx = \frac{1}{2}\int_{\partial A} \begin{bmatrix}
      x\\y
  \end{bmatrix} \cdot \mathbf n \;ds. 
\end{equation}
For a cross-section bounded by $n$ straight lines which connect the vertices $P_i=(x_i,y_i)$, $i=1,...,n$, we have 
\begin{equation}\label{eq::area}
    A = \frac{1}{2}\sum_{i=1}^n \int_{\xi=0}^1 \left(\begin{bmatrix}
      x_i(\xi)\\y_i(\xi)
  \end{bmatrix} \cdot \frac{\mathbf R \,\mathbf t}{|\mathbf t|}\right)  |\mathbf t|\;d\xi
 = \frac{1}{2}\sum_{i=1}^n \left(\hat x_i \bar y_i - \hat y_i \bar x_i \right),
\end{equation}
with the tangent vector $\mathbf t = P_{i+1}-P_{i}$, the $90^\circ$ rotation matrix $\mathbf R$, and 
\begin{equation*}
\hat x_i = \frac{x_{i+1} + x_i}{2}, \quad   \hat y_i = \frac{y_{i+1} + y_i}{2},\quad \bar x_i = \frac{x_{i+1} - x_i}{2}, \quad   \bar y_i = \frac{y_{i+1} - y_i}{2}.    
\end{equation*}
To compute the center of gravity and the moments of inertia of area, it is necessary to evaluate integrals of the general form  $I^{kl} = \int_A x^k y^l \,dx$. Setting $V = \tfrac{1}{2(k+l+2)} [x^{k+1}y^l,x^ky^{l+1}]^\top$ allows us to transform $I^{kl}$ to a sum of boundary integrals, which are then symbolically integrated using SymPy\footnote{\url{https://www.sympy.org}, accessed 30.6.2023}.
Finally, we note that ``mechpy'' can also treat circular arcs as boundary curves. The derivation is, however, more involved and skipped here for the sake of brevity.

\section{Students' evaluation of the concept}

\begin{figure}[h]
    \begin{minipage}[b]{0.33\textwidth}	\centering 	
        \includegraphics[width=0.995\textwidth,trim=2cm 8cm 1.5cm 4cm,clip]{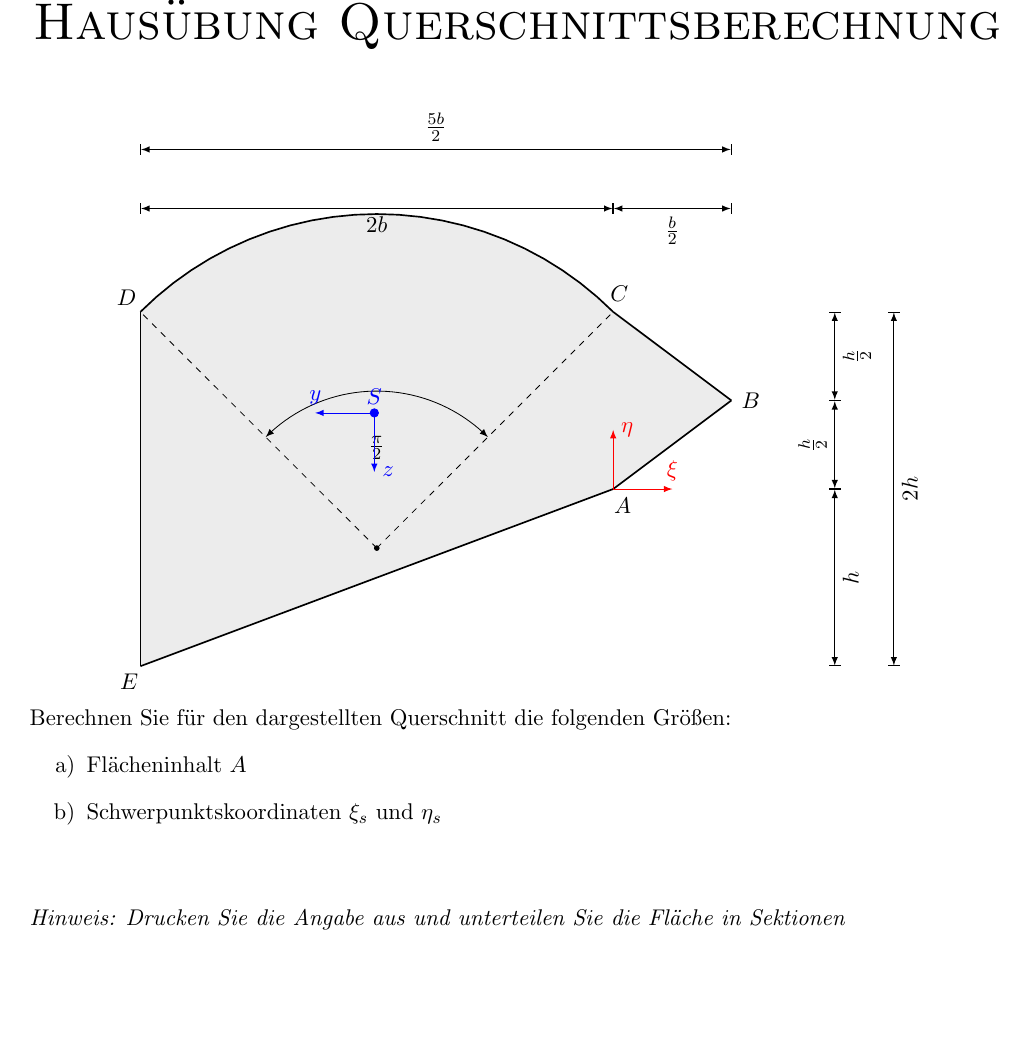}\\
        {a) Cross-section calculations}
    \end{minipage}\hfill
    \begin{minipage}[b]{0.33\textwidth}		\centering 
        \includegraphics[width=0.995\textwidth,trim=2cm 8cm 1.5cm 4cm,clip]{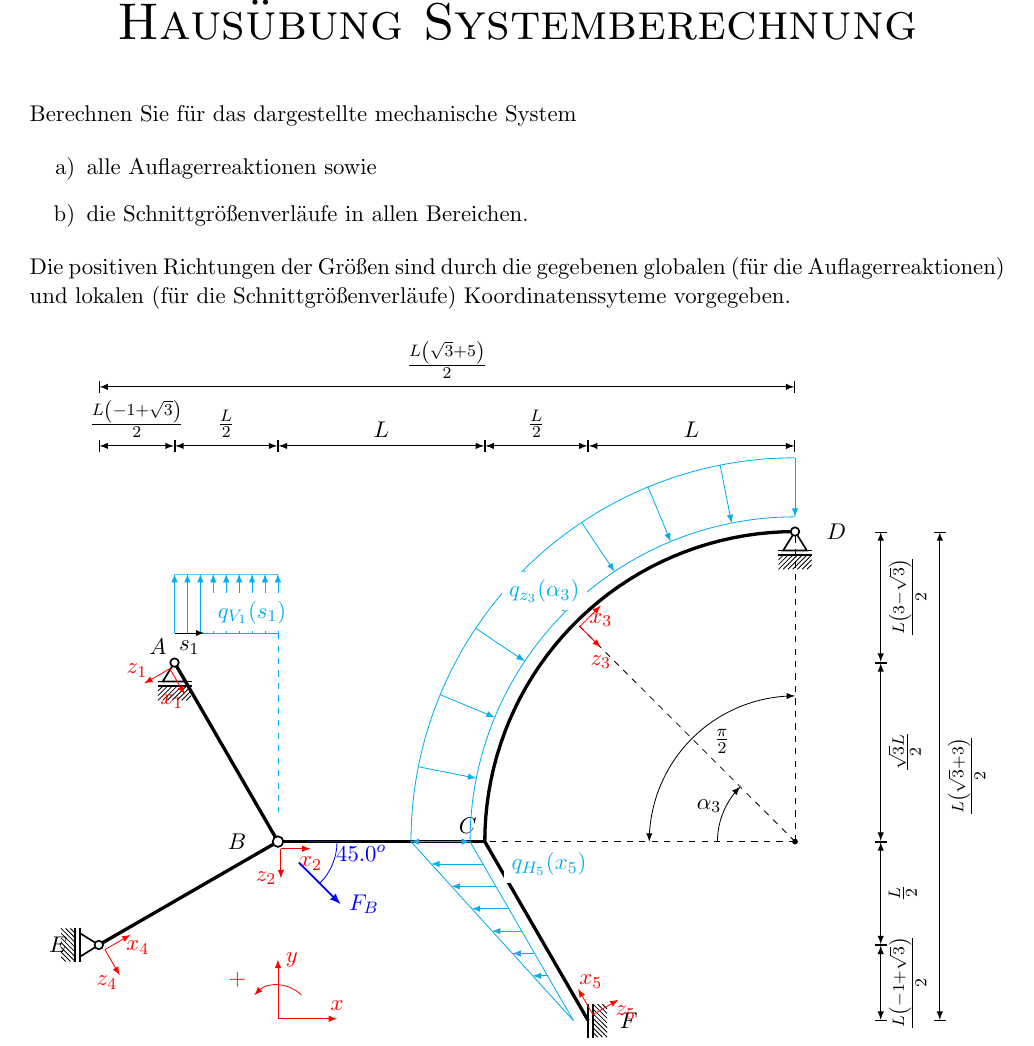}\\
        {b) Mechanical beam system}
    \end{minipage}\hfill\begin{minipage}[b]{0.33\textwidth}		\centering 
        \includegraphics[width=0.995\textwidth,trim=2cm 8cm 1.5cm 4cm,clip]{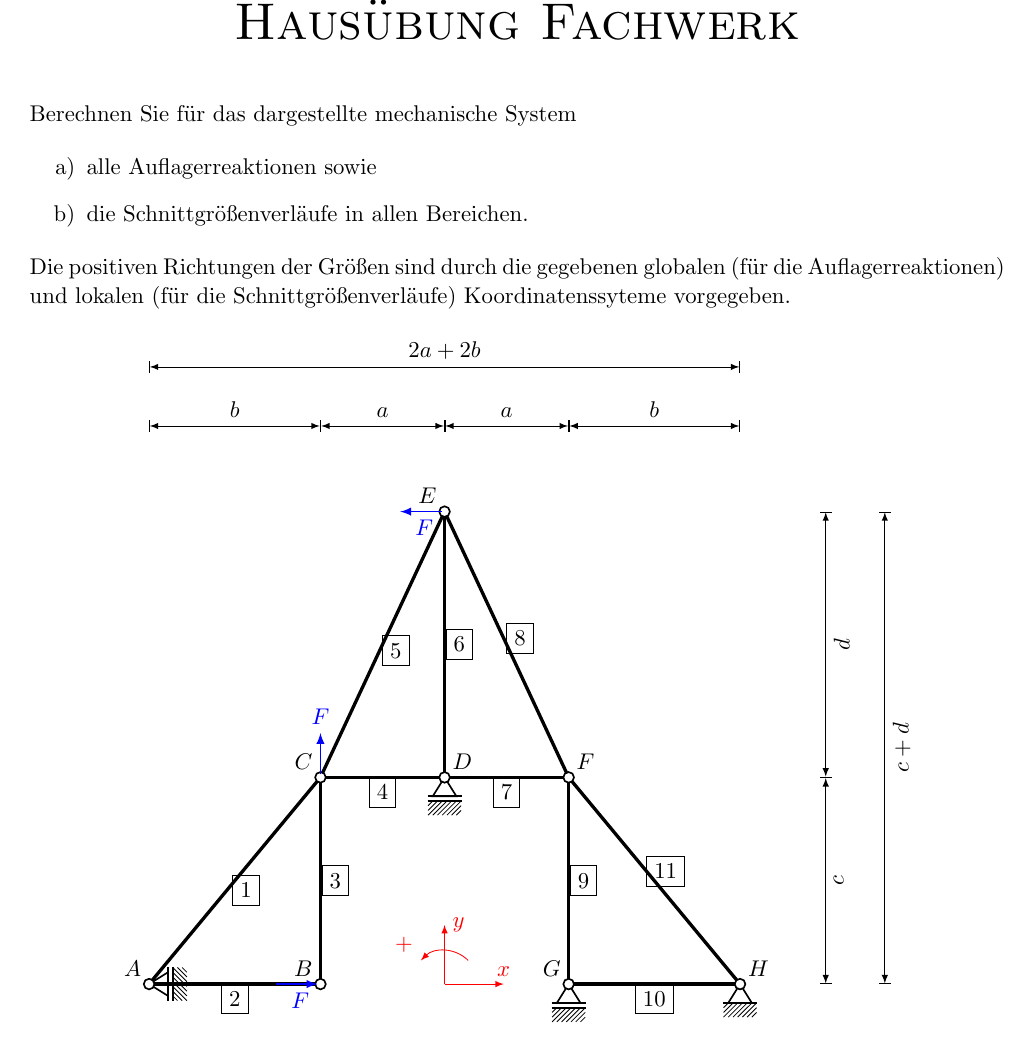}\\
        {c) Mechanical truss system}
    \end{minipage}
    \caption{Examples of the three different assignments types used in the statics lecture.}
    \label{fig:assignments}
\end{figure}

In the winter semester of 2022/23, the proposed exercise assignment technology was implemented in a lecture on statics which is part of the first semester in the bachelor's curriculum ``Bauingenieurwissenschaften und Wirtschaftsingenieurwesen''. In total, 235 students were enrolled in the course, and 143 of them completed all three exercise assignments that were part of the lecture (see \cref{fig:assignments} and \cref{tab:my_label}).  
\begin{table}[h]
    \centering
    \begin{tabular}{c c|c c c}
    \toprule
                 & topic   & total number of attempts & students & average points \\
                 \midrule
    Assignment 1 & cross-section calculations  & 534 & 173 & 2.40 / 3 \\
    Assignment 2 & mechanical beam system  & 691 & 143 & 7.29 / 10 \\
    Assignment 3 & mechanical truss system  & 531  & 166 & 3.27 / 4 \\
    \bottomrule
    \end{tabular}
    \caption{Overview of the assignments in the statics lecture detailing the total number of (re-)submissions, participation students, and points obtained. \Cref{fig:assignments} illustrates corresponding examples. }
    \label{tab:my_label}
\end{table}%

In order to assess the students' opinion and acceptance of our concept, we conducted a quantitative survey. Two surveys with standardized questionnaires were conducted, with the informed consent of the participating students.
The first questionnaire (Q1) was administered at the start of the semester, gathering data regarding the students' (i) 
demographics, (ii) learning behavior, and (iii) prior knowledge of the course's subjects. The second questionnaire (Q2) was administered at the end of the semester but before the final written exam, gathering data on (i) course satisfaction and (ii) students' opinions on the individual exercise assignments, including an adaptation of the technology-acceptance questionnaire presented in \cite{ghani2019questionnaire}.
Eighty-four students participated in Q1, and 51 students participated in Q2, with an identified overlap of 40 students between Q1 and Q2.
From the information gathered in Q1, we could see that the majority of students was male (73.8\%) and in their first semester of studies (85.7\%). Almost three-quarters of the participants (72.6\%) were 20 years old or younger. However, 44\% of them indicated having at least some prior knowledge of the subject. 
In this paper, we are mainly interested in the students' perception of and satisfaction with the new individual exercise assignment concept and, therefore, report results from Q2, with a focus on the concept's evaluation. 

\begin{figure}[h]
	\begin{minipage}[b]{0.49\textwidth}
        \centering 
        \includegraphics[scale=0.3,trim=1.7cm 2cm 2cm 1.8cm,clip]{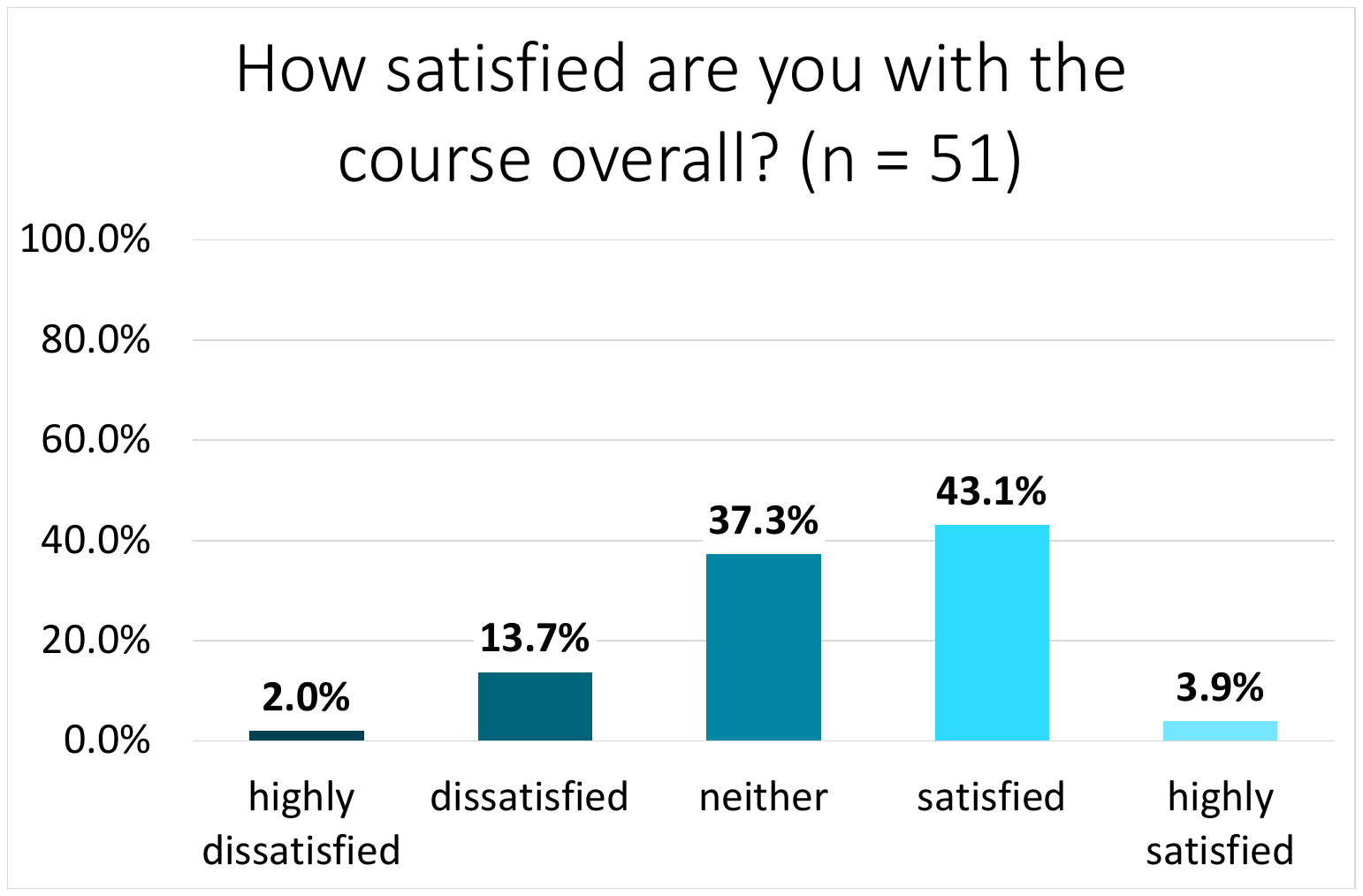}\\
        {a) Course }
    \end{minipage}    
	\begin{minipage}[b]{0.49\textwidth}
        \centering 
        \includegraphics[scale=0.3,trim=1.7cm 2cm 2cm 1.8cm,clip]{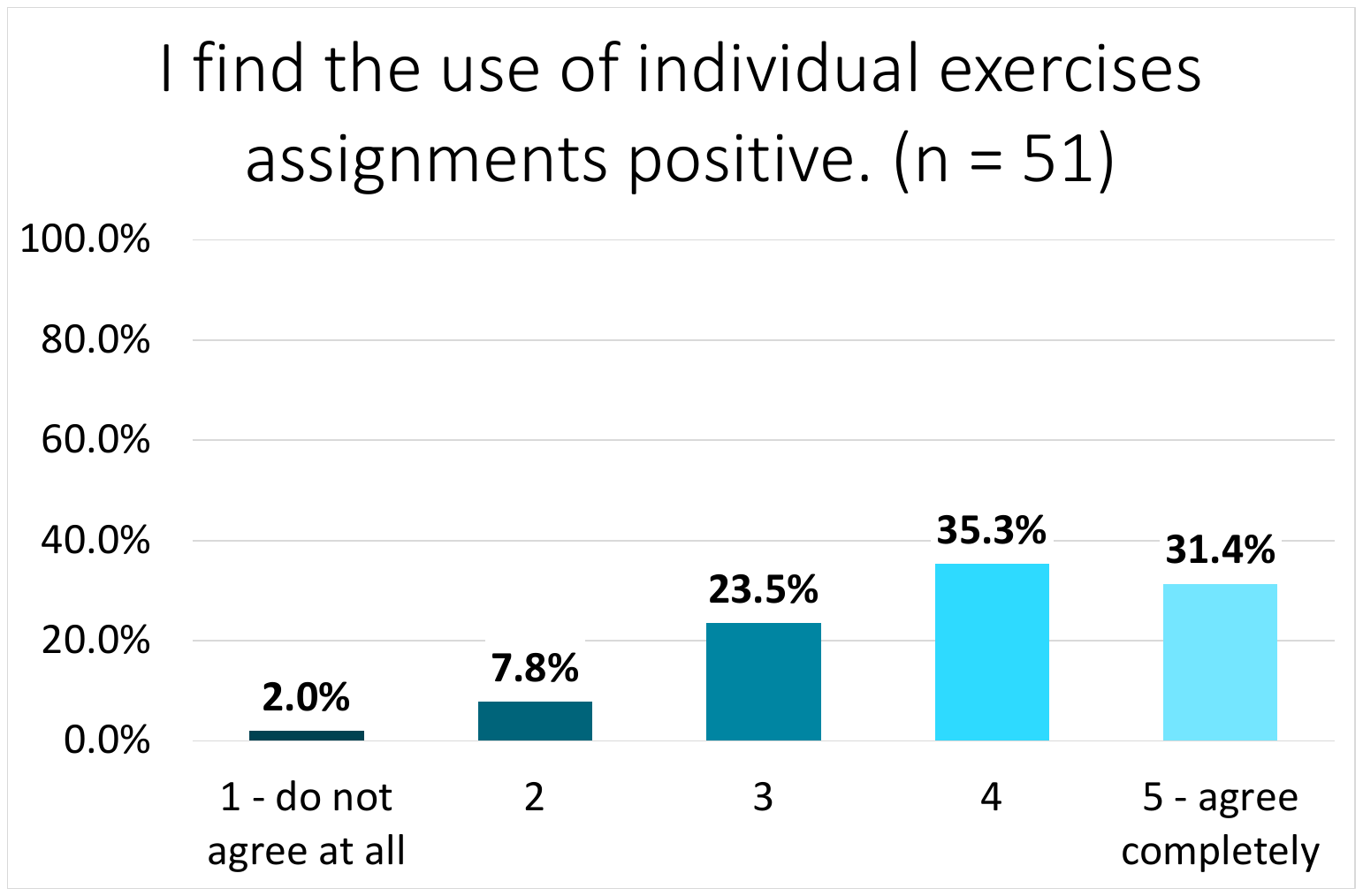}\\
        {b) Individual exercise assignments }
     \end{minipage} 
    \caption{General student feedback on (a) the course in general and (b) the individual exercise assignments in particular.} 
    \label{fig:GeneralFeedback}     
\end{figure}
\Cref{fig:GeneralFeedback} summarizes the student's general satisfaction with the course. 
The results suggest that, overall, the students participating in the survey were relatively satisfied with the course, with 47\%\ being satisfied or highly satisfied, 37.7\%\ being neither satisfied nor dissatisfied, and only 15.7\%\ being dissatisfied or highly dissatisfied (\cref{fig:GeneralFeedback}a). 
At the same time, the individual exercise assignments were received even more positively -- more than 66.7\%\ of all participants found the usage of the exercises in the course positive (\cref{fig:GeneralFeedback}b) which suggests a noticeable added value for the course.
\mb{This takeaway is also supported by the technology-acceptance results, which are clearly above scale mean (1-5 point scale, higher values indicate more positive evaluation; scale mean: 3) with $M = 3.76$ for the perceived usefulness of the individual exercise assignments (sample size $n = 51$, standard deviation $SD = 0.62$), and $M = 3.50$ for usability ($n = 51$, $SD = 0.79$). A mean of $M = 2.67$, slightly below the scale mean, was calculated in the category ease of use ($n = 51$, $SD = 0.88$),} \km{indicating that the technology is perceived as rather complicated to use and improvements might be needed in this regard.}
At the same time, it is not completely clear if this feedback relates to the use of the STACK interface or the complexity of the exercise assignments given.

When reviewing the open-text responses, two advantages of the proposed concept stood out:
\begin{enumerate}
    \item Students appreciated that the individual exercise assignments encouraged self-directed and reflective learning.
    \item Students liked that the individual exercise assignment supports learning the course content and helps identify essential learning goals.
\end{enumerate}

\begin{figure}[t]
    \centering 
    \begin{minipage}[b]{1.0\textwidth}	
    \centering \includegraphics[width=1.0\linewidth]{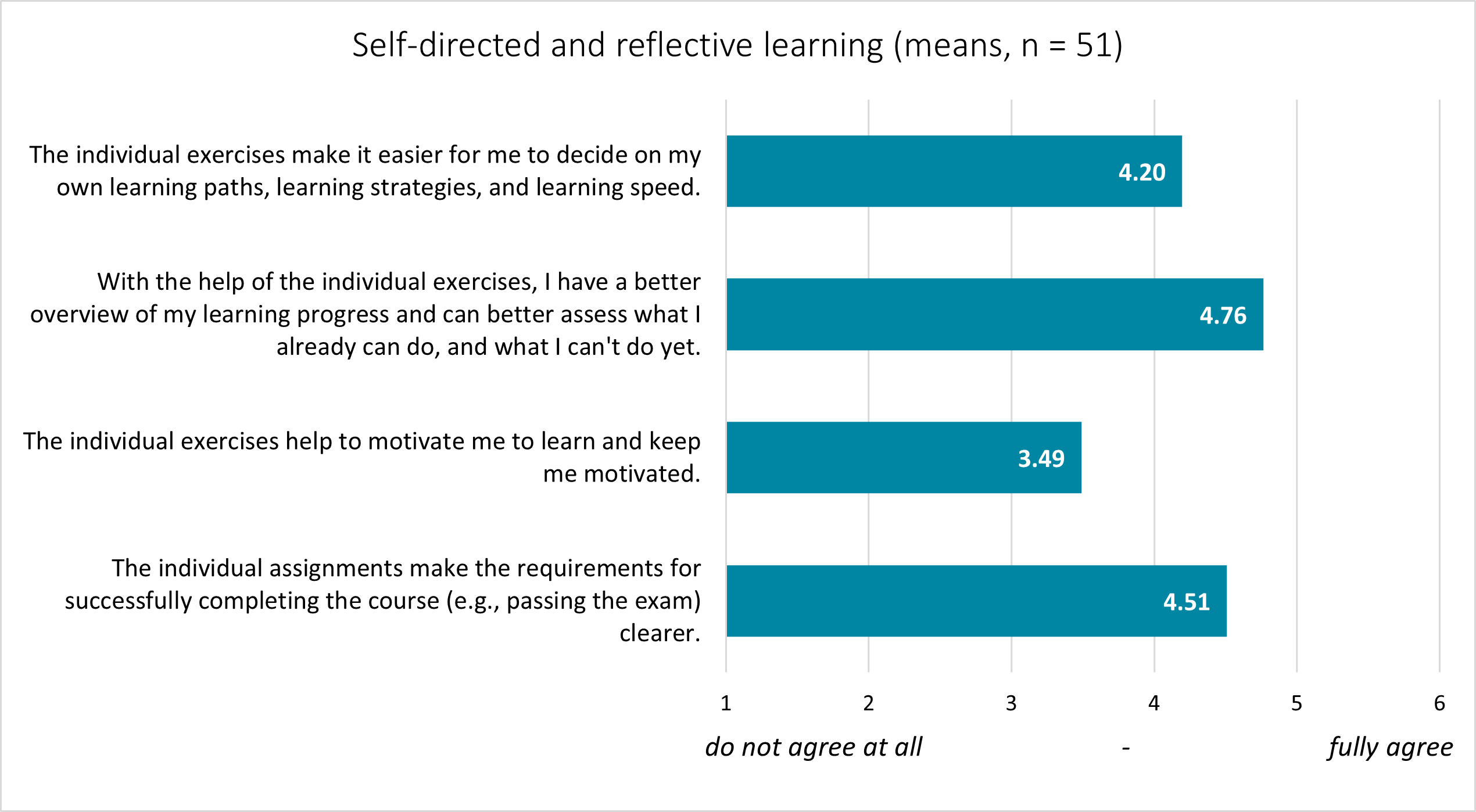}
    \end{minipage}
    \caption{Student feedback on the questions related to self-directed and reflective learning, 6-point Likert-type scale}
    \label{fig:SRL}     
\end{figure}

\km{These results are also supported by the students' answers to four questions directly related to self-regulated and reflective learning that were also part of Q2. As can be seen in~\cref{fig:SRL}, students' answers indicate that the individual exercises helped them plan and monitor their learning processes, but did not necessarily have a positive effect on their learning motivation.}

\section{Conclusions}

Based on the presented students' evaluation, we conclude that the proposed individual exercise assignment concept fosters self-directed and reflective learning. Therefore, it is an added value for the students of our mechanics lectures. 
Furthermore, using ``mechpy'' and Moodle STACK questions supports lecturers by decoupling their workload from the number of assignments.
In a follow-up paper, we will analyze the students' evaluations in more detail and draw our attention on how the concept can be further improved from a didactic point of view.

\section*{Acknowledgment}
  \mb{The development, testing, and evaluation of the individual exercise assignments were funded by the internal funding and innovation program “Digital TU Graz Marketplace – TEL Marketplace” of the Vice Rectorate for Digitization and Change Management of Graz University of Technology.}

\vspace{\baselineskip}
\bibliographystyle{myplainnat}
\bibliography{PAMM2023}

\end{document}